\documentstyle[12pt]{article}

\catcode`\@=11
\long\def\@makefntext#1{ %\parindent 1em
\protect\noindent \hbox to 3.2pt {\hskip-.9pt  
$^{{\ninerm\@thefnmark}}$\hfil}#1\hfill} %can be used 

\def\thefootnote{\fnsymbol{footnote}}
 \def\@makefnmark{\hbox to 0pt{$^{\@thefnmark}$\hss}}  %original 
        
\def\ps@myheadings{\let\@mkboth\@gobbletwo
\def\@oddhead{\hbox{} %\sl
\rightmark\hfil\ninerm\thepage}   
\def\@oddfoot{}\def\@evenhead{\ninerm\thepage\hfil %\sl
\leftmark\hbox{}}\def\@evenfoot{}
\def\sectionmark##1{}\def\subsectionmark##1{}}

\begin{document}

%----------------------------PROCSLA.STY---------------------------------------
\newcommand{\symbolfootnote}{\renewcommand{\thefootnote}
        {\fnsymbol{footnote}}}
\renewcommand{\thefootnote}{\fnsymbol{footnote}}
\newcommand{\alphfootnote}
        {\setcounter{footnote}{0}
         \renewcommand{\thefootnote}{\sevenrm\alph{footnote}}}

%------------------------------------------------------------------------------
%NEW DEFINED SECTION COMMANDS 
\newcounter{sectionc}\newcounter{subsectionc}\newcounter{subsubsectionc}
\renewcommand{\section}[1] {\vspace{0.6cm}\addtocounter{sectionc}{1} 
\setcounter{subsectionc}{0}\setcounter{subsubsectionc}{0}\noindent 
        {\bf\thesectionc. #1}\par\vspace{0.4cm}}
\renewcommand{\subsection}[1] {\vspace{0.6cm}\addtocounter{subsectionc}{1} 
        \setcounter{subsubsectionc}{0}\noindent 
        {\it\thesectionc.\thesubsectionc. #1}\par\vspace{0.4cm}}
\renewcommand{\subsubsection}[1]
{\vspace{0.6cm}\addtocounter{subsubsectionc}{1}
        \noindent {\rm\thesectionc.\thesubsectionc.\thesubsubsectionc. 
        #1}\par\vspace{0.4cm}}
\newcommand{\nonumsection}[1] {\vspace{0.6cm}\noindent{\bf #1}
        \par\vspace{0.4cm}}
                                                 
%NEW MACRO TO HANDLE APPENDICES
\newcounter{appendixc}
\newcounter{subappendixc}[appendixc]
\newcounter{subsubappendixc}[subappendixc]
\renewcommand{\thesubappendixc}{\Alph{appendixc}.\arabic{subappendixc}}
\renewcommand{\thesubsubappendixc}
        {\Alph{appendixc}.\arabic{subappendixc}.\arabic{subsubappendixc}}

\renewcommand{\appendix}[1] {\vspace{0.6cm}
        \refstepcounter{appendixc}
        \setcounter{figure}{0}
        \setcounter{table}{0}
        \setcounter{equation}{0}
        \renewcommand{\thefigure}{\Alph{appendixc}.\arabic{figure}}
        \renewcommand{\thetable}{\Alph{appendixc}.\arabic{table}}
        \renewcommand{\theappendixc}{\Alph{appendixc}}
        \renewcommand{\theequation}{\Alph{appendixc}.\arabic{equation}}
%       \noindent{\bf Appendix \theappendixc. #1}\par\vspace{0.4cm}}
        \noindent{\bf Appendix \theappendixc #1}\par\vspace{0.4cm}}
\newcommand{\subappendix}[1] {\vspace{0.6cm}
        \refstepcounter{subappendixc}
        \noindent{\bf Appendix \thesubappendixc. #1}\par\vspace{0.4cm}}
\newcommand{\subsubappendix}[1] {\vspace{0.6cm}
        \refstepcounter{subsubappendixc}
        \noindent{\it Appendix \thesubsubappendixc. #1}
        \par\vspace{0.4cm}}

%------------------------------------------------------------------------------
%MARCO FOR ABSTRACT BLOCK
\def\abstracts#1{{
        \centering{\begin{minipage}{30pc}\tenrm\baselineskip=12pt\noindent
        \centerline{\tenrm ABSTRACT}\vspace{0.3cm}
        \parindent=0pt #1
        \end{minipage} }\par}} 

%------------------------------------------------------------------------------
%NEW MACRO FOR BIBLIOGRAPHY
\newcommand{\bibit}{\it}
\newcommand{\bibbf}{\bf}
\renewenvironment{thebibliography}[1]
        {\begin{list}{\arabic{enumi}.}
        {\usecounter{enumi}\setlength{\parsep}{0pt}
%1.25cm IS STRICTLY FOR PROCSLA.TEX ONLY
\setlength{\leftmargin 1.25cm}{\rightmargin 0pt}
%0.52cm IS FOR NEW DATA FILES
%\setlength{\leftmargin 0.52cm}{\rightmargin 0pt}
         \setlength{\itemsep}{0pt} \settowidth
        {\labelwidth}{#1.}\sloppy}}{\end{list}}

%------------------------------------------------------------------------------
%FOLLOWING THREE COMMANDS ARE FOR 'LIST' COMMAND.
\topsep=0in\parsep=0in\itemsep=0in
\parindent=1.5pc

%LIST ENVIRONMENTS
\newcounter{itemlistc}
\newcounter{romanlistc}
\newcounter{alphlistc}
\newcounter{arabiclistc}
\newenvironment{itemlist}
        {\setcounter{itemlistc}{0}
         \begin{list}{$\bullet$}
        {\usecounter{itemlistc}
         \setlength{\parsep}{0pt}
         \setlength{\itemsep}{0pt}}}{\end{list}}

\newenvironment{romanlist}
        {\setcounter{romanlistc}{0}
         \begin{list}{$($\roman{romanlistc}$)$}
        {\usecounter{romanlistc}
         \setlength{\parsep}{0pt}
         \setlength{\itemsep}{0pt}}}{\end{list}}

\newenvironment{alphlist}
        {\setcounter{alphlistc}{0}
         \begin{list}{$($\alph{alphlistc}$)$}
        {\usecounter{alphlistc}
         \setlength{\parsep}{0pt}
         \setlength{\itemsep}{0pt}}}{\end{list}}

\newenvironment{arabiclist}
        {\setcounter{arabiclistc}{0}
         \begin{list}{\arabic{arabiclistc}}
        {\usecounter{arabiclistc}
         \setlength{\parsep}{0pt}
         \setlength{\itemsep}{0pt}}}{\end{list}}

%------------------------------------------------------------------------------
%FIGURE CAPTION
\newcommand{\fcaption}[1]{
        \refstepcounter{figure}
        \setbox\@tempboxa = \hbox{\tenrm Fig.~\thefigure. #1}
        \ifdim \wd\@tempboxa > 6in
           {\begin{center}
        \parbox{6in}{\tenrm\baselineskip=12pt Fig.~\thefigure. #1 }
            \end{center}}
        \else
             {\begin{center}
             {\tenrm Fig.~\thefigure. #1}
              \end{center}}
        \fi}

%TABLE CAPTION
\newcommand{\tcaption}[1]{
        \refstepcounter{table}
        \setbox\@tempboxa = \hbox{\tenrm Table~\thetable. #1}
        \ifdim \wd\@tempboxa > 6in
           {\begin{center}
        \parbox{6in}{\tenrm\baselineskip=12pt Table~\thetable. #1 }
            \end{center}}
        \else
             {\begin{center}
             {\tenrm Table~\thetable. #1}
              \end{center}}
        \fi}

%------------------------------------------------------------------------------
%ACKNOWLEDGEMENT: this portion is from John Hershberger
\def\@citex[#1]#2{\if@filesw\immediate\write\@auxout
        {\string\citation{#2}}\fi
\def\@citea{}\@cite{\@for\@citeb:=#2\do
        {\@citea\def\@citea{,}\@ifundefined
        {b@\@citeb}{{\bf ?}\@warning
        {Citation `\@citeb' on page \thepage \space undefined}}
        {\csname b@\@citeb\endcsname}}}{#1}}

\newif\if@cghi
\def\cite{\@cghitrue\@ifnextchar [{\@tempswatrue
        \@citex}{\@tempswafalse\@citex[]}}
\def\citelow{\@cghifalse\@ifnextchar [{\@tempswatrue
        \@citex}{\@tempswafalse\@citex[]}}
\def\@cite#1#2{{$\null^{#1}$\if@tempswa\typeout
        {IJCGA warning: optional citation argument 
        ignored: `#2'} \fi}}
\newcommand{\citeup}{\cite}

%------------------------------------------------------------------------------
%FOR FNSYMBOL FOOTNOTE AND ALPH{FOOTNOTE} 
\def\fnm#1{$^{\mbox{\scriptsize #1}}$}
\def\fnt#1#2{\footnotetext{\kern-.3em
        {$^{\mbox{\sevenrm #1}}$}{#2}}}

%------------------------------------------------------------------------------
\font\twelvebf=cmbx10 scaled\magstep 1
\font\twelverm=cmr10 scaled\magstep 1
\font\twelveit=cmti10 scaled\magstep 1
\font\elevenbfit=cmbxti10 scaled\magstephalf
\font\elevenbf=cmbx10 scaled\magstephalf
\font\elevenrm=cmr10 scaled\magstephalf
\font\elevenit=cmti10 scaled\magstephalf
\font\bfit=cmbxti10
\font\tenbf=cmbx10
\font\tenrm=cmr10
\font\tenit=cmti10
\font\ninebf=cmbx9
\font\ninerm=cmr9
\font\nineit=cmti9
\font\eightbf=cmbx8
\font\eightrm=cmr8
\font\eightit=cmti8

\def\mod{{\rm mod} }
%----------------------START OF DATA FILE------------------------------

\centerline{\tenbf SYMMETRIES IN NUCLEI AND MOLECULES}
%\baselineskip=16pt
%\centerline{\tenbf MANUSCRIPT USING COMPUTER SOFTWARE}
\vspace{0.8cm}
\centerline{\tenrm Dennis BONATSOS, P. KOLOKOTRONIS, D. LENIS}
\baselineskip=13pt
\centerline{\tenit Institute of Nuclear Physics, NCSR ``Demokritos''}
\baselineskip=12pt
\centerline{\tenit GR-15310 Aghia Paraskevi, Attiki, Greece}
\vspace{0.3cm}
\centerline{\tenrm C. DASKALOYANNIS, G. A. LALAZISSIS}
\baselineskip=13pt
\centerline{\tenit Department of Physics, Aristotle University of Thessaloniki}
\baselineskip=12pt 
\centerline{\tenit GR-54006 Thessaloniki, Greece}
\vspace{0.3cm}
%\centerline{\tenrm and}
%\vspace{0.3cm}
\centerline{\tenrm S. B. DRENSKA, N. MINKOV, P. P. RAYCHEV, R. P. ROUSSEV}
\baselineskip=13pt
\centerline{\tenit Institute for Nuclear Research and Nuclear Energy, 
Bulgarian Academy of Sciences}
\baselineskip=12pt 
\centerline{\tenit 72 Tzarigrad Road, BG-1784 Sofia, Bulgaria}
\vspace{0.9cm}
\abstracts{Recent progress in two different fronts is reported.
First, the concept of bisection of a harmonic oscillator or hydrogen atom, 
used in 
the past in establishing the connection between U(3) and O(4),  is 
generalized into multisection (trisection, tetrasection, etc). It is then 
shown that all symmetries of the N-dimensional anisotropic harmonic oscillator
with rational ratios of frequencies (RHO), some of which are underlying 
the structure of superdeformed and hyperdeformed nuclei, 
can be obtained from the U(N)
symmetry of the corresponding isotropic oscillator with an appropriate 
combination of multisections. Furthermore, it is seen that bisections of 
the N-dimensional hydrogen atom, which possesses an O(N+1) symmetry, lead to 
the U(N) symmetry, so that further multisections of the hydrogen atom lead 
to the symmetries of the N-dim RHO.   
 The opposite is in general not true, i.e. multisections of U(N) do 
not lead to O(N+1) symmetries, the only exception
being the occurence of O(4) after the bisection of U(3). 
Second, it is shown that there is evidence that
the recently observed in superdeformed nuclear bands 
$\Delta I=4$ bifurcation is also occuring in 
rotational bands of diatomic molecules. In addition there is 
evidence that a $\Delta I=8$ bifurcation, of the same order 
of magnitude as the $\Delta I=4$ one, is observed in superdeformed 
nuclear bands and rotational bands of diatomic molecules. 
}

\vfil
%\vspace{0.8cm}
\rm\baselineskip=14pt

\section{Introduction}
In this paper two different topics will be presented. The first (sections
2--9) is related to the symmetries of the anisotropic harmonic oscillator 
with rational ratios of frequencies. The second (sections 10--14) is related 
to a novel staggering effect observed recently in superdeformed nuclear 
bands.  

\section{Multisections of the harmonic oscillator}

Anisotropic harmonic oscillators with rational ratios of frequencies (RHOs)
%\cite{JM,Dem,Dui,Mai,Ven,MV,Cis} 
$^{1-7}$ are 
of current interest in several branches of physics. Their symmetries form
the basis for the understanding 
% \cite{Mot,Rae,Ros,Bha,Naz} 
$^{8-12}$ of the occurence of 
superdeformed and hyperdeformed nuclear shapes \cite{Nol,Jan} at very high 
angular momenta. In addition, 
 they have been recently connected \cite{RZ,ZR} to the underlying geometrical 
structure in the Bloch--Brink $\alpha$-cluster model \cite{Bri}. They are also
becoming of interest for the interpretation of the observed shell 
structure in atomic clusters \cite{Mar}, especially after the realization that 
large deformations can occur in such systems \cite{Bul}.  
An interesting problem is to what extend the various symmetries of the RHOs,
occuring for different frequency ratios, are related to other known 
symmetries. A well-known example is the case of the 3-dimensional RHO 
with frequency ratios 2:2:1, which is known to possess the O(4) symmetry
\cite{Rav}. 

In this study we show how the symmetries of the N-dim RHO can be obtained 
from the U(N) symmetry of the corresponding isotropic harmonic oscillator 
(HO) by appropriate symmetry operations, namely multisections, which are 
generalizations of the concept of bisection, introduced in \cite{Rav}. It will 
furthermore 
be shown that these symmetries can also be obtained from the O(N+1) 
symmetry of the N-dim hydrogen atom, since a bisection leads from O(N+1)
to U(N), so that further multisections lead to RHO symmetries. 
However, despite the fact that the N-dim RHO symmetries can be 
obtained from the O(N+1) symmetry by appropriate multisections, they are
not orthogonal symmetries themselves (with the exception of 2:2:1
mentioned above). 

In section 3 of this paper the Hamiltonian of the N-dim RHO is given. 
Multisections of the 3-dim and  the N-dim harmonic oscillator are
defined and used in obtaining the symmetries of the various RHOs 
in sections 4 and 5 respectively. The N-dim hydrogen atom is briefly
described in section 6. A similar
procedure is followed in sections 7 and 8 
 for the 3-dim and the N-dim hydrogen atom, while section 9
contains a summary of the present results and implications for further 
work. 

\section{ The anisotropic  harmonic oscillator with rational ratios of 
frequencies (RHO)} 

The Hamiltonian of the N-dim RHO reads  
\begin{equation}
 H = {1 \over 2} \sum_{k=1}^N \left( p_k^2 + {x_k^2\over m_k^2}\right),
\end{equation}
where $m_i$ are natural numbers prime to each other. The energy eigenvalues
are given by 
\begin{equation}
E = \sum_{k=1}^N {1\over m_k} \left( n_k+{1\over 2}\right),
\end{equation}
where $n_k$ is the number of quanta in the $k$-th direction.

\section{ Multisections of the 3-dimensional oscillator}

Let us consider the completely symmetric irreps of U(3), $[N00]$, the 
dimensions of which are given by 
\begin{equation}
d(N) = {(N+1)(N+2)\over 2}, \qquad N=0, 1, 2, \ldots 
\end{equation}  
Using the Cartesian notation  
$(n_x, n_y, n_z)$ for the U(3) states, as in \cite{Rav}, 
one can easily see that a {\bf bisection} of the 1:1:1 RHO states
(having degeneracies 1, 3, 6, 10, 15, 21, \dots, given by eq. (3)) 
distinguishing 
states with $\mod(n_z,2)=0$ and states with $\mod(n_z,2)=1$,  results it two
interleaving 2:2:1 sets of levels, having degeneracies 1, 2, 4, 6, 9, 12, 
16, 20, \dots. 

By analogy, a {\bf trisection} can be made  by distinguishing states 
with $\mod(n_z,3)=0$ or $\mod(n_z,3)=1$ or $\mod(n_z,3)=2$.
The degeneracies obtained are 1, 2, 3, 5, 7, 9, 12, 15, 18, \dots, which 
correspond 
to the 3:3:1 RHO. Therefore a trisection of the 1:1:1  HO results in three
interleaving sets of 3:3:1 RHO states. 

Similarly a {\bf tetrasection} is defined by selecting states with 
$\mod(n_z,4)=0$,  or 1, or 2, or 3. 
The degeneracies obtained are 1, 2, 3, 4, 6, 8, 10, 12, 15, 18, \dots,
which characterize the 4:4:1 RHO.   Therefore a tetrasection of the 
1:1:1 HO leads to four interleaving sets of 4:4:1 RHO states. 

In general, an {\sl n-section} of the 1:1:1 HO is obtained by separating 
states with $mod(n_z, n)$ =0, or 1, or 2, \dots, or $n-1$. In this case 
$n$ interleaving sets of the n:n:1 RHO states, which corresponds to an 
oblate shape,  are obtained. 
It is clear that n-sections using $n_x$ or $n_y$ instead of $n_z$ lead 
to the same conclusions. 

One can consider successively more than one bisections, trisections, etc. 
Let us consider more than one bisections first. 

Getting the results of the $\mod(n_z,2)=0$ bisection of the HO and 
applying a $\mod(n_y,2)$ =0 bisection on them we obtain 
the degeneracy pattern 1, 1, 3, 3, 6, 6, 10, 10, i.e. ``two copies''
of the 1:1:1 degeneracies,  which corresponds 
to the 2:1:1 RHO. The same result is obtained for any combination of 
two bisections along two differerent axes. 

Bisecting the 1:1:1 HO for a third time, along the $x$-axis this time by
using $\mod(n_x,2)$ =0, one obtains
the degeneracy pattern 1, 3, 6, 10, \dots, i.e. that of the original
1:1:1 HO. 

Furthermore one can easily see that:

i) Two trisections along different axes lead to degeneracies 1, 1, 1, 3, 3, 3, 
6, 6, 6, \dots, i.e. to the 3:1:1 RHO pattern (``three copies'' of the 
1:1:1 degeneracies).  

ii) Three trisections lead to the original 1:1:1 HO degeneracy pattern. 

iii) Two tetrasections lead to degeneracies 1, 1, 1, 1, 3, 3, 3, 3, 6, 6, 6, 6,
\dots, i.e. to the 4:1:1 RHO pattern (``four copies'' of the 1:1:1 
degeneracies).  

iv) Three tetrasections lead back to the original 1:1:1 HO pattern. 

The results obtained so far are summarized in Table 1. 

\begin{table}
\centerline{\bf Table 1}

Degeneracies of various 3-dim anisotropic harmonic oscillators with 
rational ratios of frequencies (RHOs)  obtained from the U(3) symmetry of the 
isotropic 3-dim
harmonic oscillator (HO) by the application of various multisections. 
The first line corresponds to the isotropic 3-dim HO. In the rest of 
the lines the first column contains the appropriate multisection, while the 
second column contains the frequency ratios $m_1:m_2:m_3$ of the resulting 
RHO. \bigskip
\hrule
$$\vbox{\halign{\hfil #\hfil &&\quad \hfil #\hfil \cr
% $m_1:m_2:m_3$  & & & & & & & & & & & & & & & &   \cr
%                & & & & & & & & & & & & & & & &   \cr
 U(3)  &  1:1:1 & 1 & 3 & 6 & 10 & 15 & 21 & 28 & 36 &  & & & \cr
1 bisection &  
2:2:1 & 1 & 2 & 4 & 6 & 9 & 12 & 16 & 20 & & & & \cr
1 trisection &
3:3:1 & 1 & 2 & 3 & 5 & 7 & 9 & 12 & 15 & 18 & 22 & 26 & 30 \cr
1 tetrasection & 
4:4:1 & 1 & 2 & 3 & 4 & 6 & 8 & 10 & 12 & 15 & 18 & 21 & 24 \cr 
2 bisections & 
1:1:2 & 1 & 1 & 3 & 3 & 6 & 6 & 10 & 10 & 15 & 15 & 21 & 21 \cr
2 trisections & 
1:1:3 & 1 & 1 & 1 & 3 & 3 & 3 & 6 & 6 & 6 & 10 & 10 & 10 \cr
2 tetrasections & 
1:1:4 & 1 & 1 & 1 & 1 & 3 & 3 & 3 & 3 & 6 & 6 & 6 & 6 \cr
3 bisections & 1:1:1 & 1 & 3 & 6 & 10 & 15 & 21 & 28 & 36 & & & & \cr
3 trisections & 1:1:1 & 1 & 3 & 6 & 10 & 15 & 21 & 28 & 36 & & & & \cr
3 tetrasection & 1:1:1 & 1 & 3 & 6 & 10 & 15 & 21 & 28 & 36 & & & & \cr
}}$$
\hrule
\end{table}
In general one can see that: 

i) Two n-sections (along different axes) lead to the degeneracy pattern of 
n:1:1, i.e. to ``n copies'' of the 1:1:1 degeneracies. n:1:1 corresponds 
to a prolate shape.  

ii) Three n-sections lead back to the degeneracy pattern of the 1:1:1 HO. 

One can, of course, apply successive n-sections with different n. For
example, applying $\mod(n_z,2)=0$, $\mod(n_y,3)=0$ and $\mod(n_x,3)=0$
one obtains the degeneracy pattern 1, 1, 2, 1, 2, 4, 2, 4, 6, \dots, 
which corresponds to the 2:2:3 oscillator.  

In general one can see that by applying a $k$-section, an $l$-section and 
an $m$-section along different axes one obtains the degeneracy pattern 
$(kl):(mk):(lm)$, where common factors appearing in all three quantities 
$(kl)$, $(mk)$, $(lm)$ can be dropped out.  

We have therefore seen that all the symmetries of the 3-dim RHO can be 
obtained 
from the U(3) symmetry of the isotropic 3-dim HO by an appropriate set of 
n-sections. 

A special remark can be made about the 2:2:1 case. The degeneracies obtained 
there correspond to the dimensions of the irreps of O(4), given by 
\begin{equation} 
d(\mu_1, \mu_2) = (\mu_1+\mu_2+1) (\mu_1-\mu_2+1).
\end{equation} 
In particular, the degeneracies 1, 4, 9, 16, \dots correspond to the 
integer irreps $(\mu,0)$ with $\mu=0$, 1, 2, 3, \dots, while the degeneracies 
2, 6, 12, 20, \dots correspond to the spinor irreps $({n\over 2}, {1\over 2})$
with $n=1$, 3, 5, 7, \dots. This result has been first found by Ravenhall
{\it et al.} \cite{Rav}. It has been pointed out that O(4) is obtained by 
imposing a reflection condition on U(3). For example, O(4) is obtained by 
 selecting the states with $n_z$=odd, a procedure which is equivalent to 
the insertion of an impenetrable barrier across the $xy$ plane. 

\newpage
\section{ Multisections of the N-dimensional oscillator}

The symmetry is U(N). If N is even, there is an Sp(N) subalgebra, if N is odd
there is no such subalgebra. 

N bisections lead back to the U(N) irreps. 

N$-$1 bisections lead to the 2:1:1:\dots:1 symmetry, i.e. to ``two copies''
 of the U(N) irreps. 

N$-$2 bisections lead to the 2:2:1:1:\dots :1 symmetry, which bears certain
similarities to O(N+1). The dimensions of the integer irreps 
are obtained correctly. The dimensions of the odd irreps differ 
by a factor of $2^{\nu-1}$, where $\nu=N/2$ for $N$ even or $\nu=(N-1)/2$ for 
$N$ odd. Therefore the 2:2:1:1:\dots :1 symmetry is not in general O(N+1).   

N$-$3 bisections lead to the 2:2:2:1:1:\dots :1 degeneracies. 

Two bisections lead to the 2:2:\dots :2:1:1 degeneracies. 

One bisection leads to the 2:2:\dots :2:2:1 degeneracies. 

Similarly

one n-section leads to the n:n:\dots :n:n:1 degeneracies, 

two n-sections lead to the n:n:\dots :n:1:1 degeneracies, 

N$-$2 n-sections lead to n:n:1:\dots :1:1, 

N$-$1 n-sections lead to n:1:1:\dots :1:1, 

N n-sections lead back to U(N). 

\section{ The hydrogen atom} 

So far we have considered multisections of the N-dim harmonic 
oscillator. We are now going to consider multisections of the hydrogen 
atom (HA) in N dimensions, which is known to be characterized by the O(N+1)
symmetry \cite{All}, which is also the symmetry characterizing a particle 
constrained to move on an (N+1)-dim hypersphere. 

\section{ Multisections of the 3-dimensional hydrogen atom}

The 3-dim hydrogen atom is known to possess the O(4) symmetry. 
We know that the irreps of O(4) are characterized by two labels $\mu_1$, 
$\mu_2$ and are denoted by $(\mu_1, \mu_2)$, while the irreps of O(3) are 
characterized by one label $\mu_1'=L $ (the usual angular momentum 
quantum number) and are denoted  by $(\mu_1')$. When making the reduction 
O(4) $\supset$ O(3), $\mu_1'$ obtains all values permitted by the condition 
$\mu_1 \geq \mu_1' \geq \mu_2$ \cite{Ham}. Furthermore, the decomposition 
O(3)$\supset$O(2) can be made, the irreps of O(2) characterized  by the 
quantum number $M=L$, $L-1$, $L-2$, \dots, $-(L-1)$, $-L$. 

We are going to consider the completely symmetric irreps of O(4), which are 
of the form $(\mu_1, 0)$. Writing down the $(LM)$ states contained in each 
O(4) irrep one finds that the dimensions of the irreps are 1, 4, 9, 16, 25, 
\dots, as expected from eq. (4), since only the integer irreps occur. 
As pointed out by Ravenhall {\it et al.} \cite{Rav}, a bisection
can be effected by inserting an impenetrable barrier through the center 
of the hydrogen atom. Only the states with $L-M$=odd remain then. 

The resulting degeneracies are again 1, 3, 6, 10, 15, 21, \dots, i.e. 
U(3) degeneracies. Therefore a {\bf bisection} of the 3-dim hydrogen atom, 
effected by choosing states with $\mod(L-M, 2)=0$ or $\mod(L-M,2)=1$,  
is leading to two interleaving sets of U(3) states, having degeneracies 
1, 3, 6, 10, 15, \dots. Choosing states 
with $\mod(L+M, 2)=0$ or 1 obviously leads to the same results. 

The fact that by bisecting O(4) one obtains U(3) has been first pointed
out by Ravenhall {\it et al.} \cite{Rav}. 
Once the U(3) symmetry of the 3-dim HO
is obtained, any further multisections on it will lead to RHO degeneracies,
as pointed out in subsec. 2.1.  

\section{ Multisections of the N-dimensional hydrogen atom}

The N-dim hydrogen atom is characterized by the O(N+1) symmetry. Only the 
completely symmetric irreps of O(N+1) occur. Using the chain 
O(N+1) $\supset$ O(N) $\supset$ \dots $\supset$ O(3) $\supset$ O(2) one can 
find the (LM) states contained in each O(N+1) irrep. Bisecting them using 
$\mod(L-M,2)=0$ or 1 one is left with the irreps of U(N). Further multisections
of the U(N) irreps lead to the appropriate symmetries of the N-dim RHO. It is
therefore clear that all  symmetries of the N-dim RHO can be 
obtained from a common parent, the O(N+1) symmetry. Thus it is not 
surprising that some of them (notably the 2:2:1:\dots:1 ones) show 
similarities to the corresponding O(N+1) symmetry. However, the only case 
in which an N-dim  RHO symmetry is identical to an O(N+1) symmetry occurs 
for N=3, for which the 2:2:1 RHO symmetry is O(4) \cite{Rav}. The rest of the 
RHO symmetries are not related to any orthogonal symmetries. 

\section{ Multisections: summary}

The concept of bisection of an N-dim isotropic harmonic oscillator with U(N) 
symmetry, introduced by Ravenhall {\it et al.} \cite{Rav}, has been 
generalized. Trisections, tetrasections, \dots,
n-sections of the N-dim isotropic harmonic oscillator have been introduced. 
They are shown to lead to the various symmetries of the anisotropic 
N-dim harmonic oscillator with rational ratios of frequencies (RHO).
Furthermore, multisections of the N-dim hydrogen atom with O(N+1) symmetry
have been considered. It is shown that a bisection of O(N+1) leads to U(N),
so that further multisections just lead to various cases of the N-dim RHO.   
The opposite does not hold, i.e. multisections of U(N) do not lead to
O(N+1) symmetries, the only exception being the bisection of U(3) which
does lead to O(4). Even in the case of the 4-dim HO, which has the U(4) 
symmetry, which is isomorphic to O(6) and has an O(5) subalgebra, no 
multisection, or combination of multisections, leading to a RHO with O(5)
symmetry can  be  found. We conclude therefore that the rich variety of
the N-dim RHO symmetries have a common ``parent'', the U(N) symmetry of the 
N-dim isotropic harmonic oscillator or the O(N+1) symmetry of the N-dim 
hydrogen atom, but they are not in general related to unitary or orthogonal
symmetries themselves.  

Since the RHO is of current interest in relation to various physical 
systems (superdeformed and hyperdeformed nuclei 
%\cite{Mot,Rae,Ros,Bha,Naz},
$^{8-12}$,  
Bloch--Brink $\alpha$-cluster model 
%\cite{RZ,ZR,Bri},
$^{15-17}$,  
deformed atomic clusters \cite{Mar,Bul}), 
the unification of the rich variety 
of symmetries appearing in the RHO for different frequency ratios 
in a common algebraic framework is an interesting project. In \cite{MV}
the 3-dim RHO degeneracies  are obtained as reducible representations 
of U(3). It could be possible to construct an algebraic framework in 
which the RHO degeneracies occur as irreducible representations of an 
appropriate algebra. Work in this direction is in progress 
\cite{BDKL,Patras94}. 

Throughout this study the properties of the completely symmetric irreps 
of U(N) and O(N+1) have been considered. Similar studies of completely 
antisymmetric irreps, or irreps with mixed symmetry, might be worth exploring.

\section{ Novel staggering effects in rotational bands}

Rotational bands of diatomic molecules \cite{Herzberg} and rotational bands 
of deformed nuclei \cite{BM} have many features in common, despite the 
different energy scales involved in each case. Molecular rotational bands 
are in general closer to the behavior of the rigid rotator than their nuclear
counterparts. In the last decade much interest has been attracted by 
superdeformed nuclear bands \cite{Nol,Jan,Twin},
which are characterized by relatively high angular momenta and behavior 
closer to the rigid rotator limit in comparison to normal deformed nuclear 
bands. 

A rather surprising feature has been recently discovered 
\cite{Fli,Ced} in superdeformed 
nuclear bands: Sequences of states differing by four units of angular 
momentum are displaced relative to each other, the relative shift being 
of order of $10^{-4}$ of the energies separating the levels of these bands.  
A few theoretical proposals for the possible explanation of this 
$\Delta I=4$ bifurcation, 
 which is also called $\Delta I=2$ staggering, have
already  been made $^{30-33}$. 
%\cite{HMott,Macc,PavFli,MQ}  

A reasonable question is therefore whether $\Delta I=4$
bifurcations (i.e. $\Delta I=2$ staggering), observed in 
nuclear superdeformed bands as discussed in section 11,  also occur 
in rotational spectra of diatomic molecules. 
We are going to show in section 12
that this is indeed the case. Bifurcations with $\Delta I=8$ and 
$\Delta I=12$ will also be discussed in section 13, while section 14 
will contain a summary of the present results and discussion of further
work.  

\section{$\Delta I=2$ staggering in superdeformed nuclear bands}

In nuclear physics the experimentally determined quantities are the 
$\gamma$-ray transition energies between levels differing by two units 
of angular momentum ($\Delta I=2$). For these the symbol
\begin{equation}
 E_{2,\gamma}(I) = E(I+2)-E(I)
\label{eq:1old}
\end{equation}
is used, where $E(I)$ denotes the energy of the level with angular momentum 
$I$. 
The deviation of the $\gamma$-ray transition energies from the 
rigid rotator behavior can be measured by the quantity
\cite{Ced}
\begin{equation}
 \Delta E_{2,\gamma}(I) = {1\over 16} (6E_{2,\gamma}(I) -4E_{2,\gamma} (I-2)
-4E_{2,\gamma}(I+2) +E_{2,\gamma}(I-4) +E_{2,\gamma}(I+4)).
\end{equation}
\noindent Using the rigid rotator expression $E(I)=A I(I+1)$ one
can easily see that
in this case $\Delta E_{2,\gamma} (I) $ vanishes.  
This is due to the fact that eq. (6) is the 
discrete approximation of the fourth derivative of the function
$E_{2,\gamma}(I)$.

Several nuclear superdeformed rotational bands such as (a) to
(e) for  $^{149}$Gd 
\cite{Fli} and the bands (1) to (3) for $^{194}$Hg \cite{Ced} 
were analyzed. The corresponding tables are not included in
this short presentation,  
being reserved for a forthcoming longer publication. 
The analysis shows that the $\Delta E_{2,\gamma} (I)$ values
exhibit an anomalous staggering. It should be noted, however,
that only for the band (a) 
of $^{149}$Gd \cite{Fli}, the amplitude of the oscillations
(see for example fig. 3 of ref \cite{Ced}) is definitely
outside the experimental errorbars. The following observations can be made:

\noindent i) $\Delta E_{2,\gamma} (I)$ obtains alternating
positive and  negative 
values. This is why this effect has also been called ``$\Delta I=2$ 
staggering''. 

\noindent ii) The magnitude of $\Delta E_{2,\gamma}(I)$ is of
order $10^{-4}$--$10^{-5}$
of that of the $\gamma$-ray transition energies. 

\noindent iii) The staggering oscillation width is an increasing
function  of the 
angular momentum $I$.

\section{$\Delta I=2$ staggering in rotational bands of diatomic molecules}

In the case of molecules the experimentally determined 
quantities regard the R branch ($I\rightarrow I+1$) and the P branch 
($I\rightarrow I-1$). They are related to transition energies through the 
equations \cite{Barrow}
\begin{equation}
E^R(I)-E^P(I)= E_{v=1} (I+1) -E_{v=1} (I-1) = DE_{2, v=1} (I-1),
\end{equation}
\begin{equation}
E^R(I)-E^P(I+2) = E_{v=0}(I+2)-E_{v=0}(I)=DE_{2, v=0}(I),
\end{equation}
where in general 
\begin{equation}
 DE_2 (I) = E(I+2)-E(I),
\end{equation}
and $v$ is the vibrational quantum number. $\Delta I=2$ staggering  can then
be estimated by using eq. (6), with $E_{2,\gamma}(I)$ replaced by $DE_2(I)$:
\begin{equation}
 \Delta E_2 (I)= {1\over 16} (6 DE_2(I)-4 DE_2(I-2)-4 DE_2(I+2) +DE_2(I-4)
+DE_2(I+4)).
\end{equation}
It is noted, that for the sake of simplicity a normalized form
of the discrete fourth derivative is used in (10) as well as in
the subsequent equations : (11), (13), and (15).  
We have analyzed quite a few molecular rotational bands
for several diatomic molecules. The detailed results are saved for a longer 
publications. Some of the bands revealing a staggering effiect are:
\hfill\break
i) The (1-1), $v$=1
rotational band of the   C$^1\Sigma^+$--X$^1\Sigma^+$ system of YD 
(data from \cite{RamYD}). 
\hfill\break
ii) The (2-2), $v=$1  rotational band of the 
A$^1\Sigma^+$--X$^1\Sigma^+$ system of YN (data from \cite{RamYN}).
\hfill\break
iii) The (4-3), $v$=1 band for the molecule CS (data from \cite{RamCS}).
\hfill\break 
Similar results have been obtained from the other available bands of these
molecules and from the 
A$^6\Sigma^+$--X$^6\Sigma^+$ system of CrD (data from \cite{RamCrD}).  
The following comments are in place:

\noindent i) $\Delta E_2(I)$ exhibits alternating signs with
increasing $I$,  a 
fingerprint of $\Delta I=2$ staggering for the $v=1$ band, while
the $v=0$ band data do not permit a rigorous identification of the staggering
effect.  

\noindent ii) The magnitude of the perturbation, $\Delta E_2(I)$, is of order
$10^{-3}$--$10^{-5}$ of that of the interlevel separation energy.  

\noindent iii) The staggering oscillation width is not a
monotonically  increasing
function of the angular momentum $I$.
The irregularities in the magnitude of $\Delta E_2(I)$ might indicate the 
presence of subsequent bandcrossings \cite{MRM}. 
It is known that the bandcrossing effect is seeing only when the interaction 
between the two bands which cross each other is relatively weak
\cite{VDS}. Therefore only the levels neighboring the crossing are affected 
by the interaction. From eq. (10) it is then clear that perturbing an 
energy level results in perturbing 5 consequent values of 
$\Delta E_2(I)$. In view of this, case i)  looks very much like depicting 
two subsequent bandcrossings. Cases ii) and iii), however, do not immediately
accept such an interpretation. Bandcrossing has been recently suggested as
a possible source of the $\Delta I=4$ bifurcation in nuclei \cite{Sun,Reviol}.
Certainly more work is needed in this direction.  

\noindent iv) The staggering effect is more prominent in the
case of  even angular
momentum data than in the case of odd angular momentum data.

\section{$\Delta I=4$ and $\Delta I=6$ staggering in nuclei and molecules}

One might further wonder if bifurcations with $\Delta I >4$ can also occur. 
In the nuclear case, the existence of $\Delta I =4$ staggering can be checked
by using the quantity
\begin{equation}
 \Delta E_{4,\gamma} (I)= {1\over 16} (6 E_{4,\gamma}(I) -4 E_{4,\gamma}(I-4)
-4 E_{4,\gamma}(I+4) +E_{4,\gamma}(I-8) +E_{4,\gamma}(I+8)),
\end{equation}
where
\begin{equation}
 E_{4,\gamma}(I)= E(I+4)-E(I).
\end{equation}
\noindent Results for several superdeformed nuclear bands have
been  calculated.  A good example is provided by 
the $\Delta I=8$ bifurcation for the superdeformed 
band (a) of $^{149}$Gd \cite{Fli}. Note that no angular momentum 
assignments can be made, since they are still uncertain.
The following remarks apply:

\noindent i) $\Delta E_{4,\gamma} (I)$ acquires alternating
signs with  increasing $I$,
indicating the existence of a $\Delta I=8$ bifurcation. 

\noindent ii) The order of magnitude of the $\Delta I=4$
staggering is the same as 
that of the $\Delta I=2$ staggering.

In the case of diatomic molecules one can search for $\Delta I =4$ 
staggering by using the quantity
\begin{equation}
 \Delta E_4 (I) = {1\over 16} (6 DE_4(I) -4 DE_4(I-4) -4 DE_4(I+4)
+DE_4(I-8) +DE_4(I+8)),
\end{equation} 
where 
\begin{equation}
 DE_4 (I)= E(I+4)-E(I).
\end{equation}
\noindent In our study we have analyzed the larger known bands of the 
molecule CS, i.e. the bands (1-0) $v=1$, (2-1) $v=0$, (4-3) $v=0$, 
(2-1) $v=0$ \cite{RamCS}. 
The results
for the rotational bands of the 
A$^6\Sigma^+$--X$^6\Sigma^+$ system of CrD (data from \cite{RamCrD}),
the C$^1\Sigma^+$--X$^1\Sigma^+$ system of YD  
(data from \cite{RamYD}), 
and the A$^1\Sigma^+$--X$^1\Sigma^+$ system of YN 
(data from \cite{RamYN}) were also
considered. For these molecules  experimental data of long enough
bands exist, permitting the calculations. 
A good example is provided by the $\Delta I=4$ staggering ($\Delta I =8$
bifurcation)  for the (1-1),  $v=1$ band of the 
C$^1\Sigma^+$--X$^1\Sigma^+$ system of the  molecule YD.
The following comments can be made:

\noindent i) Alternating signs of $\Delta E_4(I)$, 
a fingerprint of $\Delta I=8$ bifurcation, are observed.

\noindent ii) The magnitude of the $\Delta I =4$ staggering appears to be 
the same as that of the $\Delta I=2$ staggering. 

$\Delta I=12$ bifurcation i.e. $\Delta I=6$ staggering can be
searched for  through use of the quantity
\begin{equation}
 \Delta E_6 (I)={1\over 16} (6DE_6(I)-4 DE_6(I-6)-4DE_6(I+6)+DE_6(I-12)
+DE_6(I+12)),
\end{equation}
where
\begin{equation}
 DE_6(I)=E(I+6)-E(I).
\end{equation}
 Calculations have been carried out for a few cases of  rotational 
 bands of CS (data from \cite{RamCS}), 
 and for the B$^1\Sigma_u^+$--X$^1\Sigma_g^+$
 system of $^{63}$Cu$^{65}$Cu (data from \cite{RamCu65}),
 in which bands long enough for such
a calculation are known. These results look like being in favor of the 
existence of $\Delta I=6$ staggering of the 
same order of magnitude as $\Delta I=4$ and  $\Delta I=2$
staggering, but they are not enough for drawing any final conclusions. 

\section{Novel staggering effects: summary}

The observation of $\Delta I=2$ and $\Delta I=4$ staggering in rotational 
spectra of diatomic molecules offers a corroboration of the existence of the
same effect in nuclei, since the experimental techniques used in each case 
are quite different, so that the occurrence of the same systematic errors 
in both cases is improbable. Furthermore, the energy scales involved in 
nuclei and molecules are very different (the separation of energy levels 
in molecules is of the order of 10$^{-2}$eV, while in nuclei of the order 
of 10$^5$eV), but the staggering effects are 
in both cases of the same order of magnitude relative to 
the separation of the 
energy levels, indicating that the same basic 
mechanism, possibly related to some perturbations of given symmetry, 
might be responsible for these effects in both cases. 

\noindent In conclusion, we have shown that: 

\noindent i) $\Delta I=2$ staggering, first observed in superdeformed
nuclear bands \cite{Fli,Ced},  
occurs as well in rotational bands of diatomic molecules. 

\noindent ii) In all cases the magnitude of the $\Delta I=2$ staggering is 
$10^{-3}$--$10^{-5}$ of that of the separation of the energy levels. 

\noindent iii) Furthermore $\Delta I=4$ staggering appears to be present both
 in superdeformed 
nuclear bands as well as in rotational bands of diatomic molecules, its 
order of magnitude being the same as that of the $\Delta I=2$ staggering
in the same physical system.

\noindent iv) 
In most cases the magnitude of the staggering does not show a simple 
dependence on angular momentum. In several cases one sees about
5 points deviating
very much for the smooth rotational behavior, then several points much 
closer to the pure rotational behavior, then again about 5
points deviating  very
much from the smooth rotational behavior, and so on. Such a picture raises
suspicions for the presence of bandcrossings at the points at which the 
large deviations occur \cite{Sun,Reviol}. 

Concerning the theoretical explanation of the $\Delta I=2$ staggering effect,
some proposals in the nuclear physics framework already exist $^{30-33}$.
% \cite{HMott,Macc,PavFli,MQ}. 
However, one cannot draw any firm 
conclusions up to now. 
In view of the present results, further efforts, 
in investigating other molecular and nuclear data, are necessary.  

\bigskip\medskip
\section { Acknowledgements}

Two authors (DB and GAL) have been supported by the E.U. under contracts
ERBCHBGCT930467 and ERBFMBICT950216 respectively. An\-other author  
(PPR) acknowledges support from the Bulgarian Ministry of Science 
and Education under contracts $\Phi$-415 and $\Phi$-547.  
Three authors (DB, CD, GAL) have been supported by the Greek Secretariat of 
Research and Technology under contract PENED95/1981.

\end{document}